\documentclass[11pt]{article}
\pdfoutput=1

\usepackage{amsmath, amsfonts, amssymb}
\usepackage{comment}
\usepackage{graphicx}
\usepackage{psfrag}
\usepackage[usenames,dvipsnames,svgnames,table]{xcolor}
\usepackage{enumerate}
\usepackage{soul}
\usepackage{subfig}
\usepackage{cite}
  \usepackage{a4wide}
  
  \usepackage{color}
  \definecolor{dark-gray}{gray}{0.20}
  \definecolor{gray}{gray}{0.30}
  \definecolor{light-gray}{gray}{0.80}
  \definecolor{dark-red}{rgb}{0.7,0,0}
  \definecolor{dark-green}{rgb}{0.1,0.4,0}
  \definecolor{dark-blue}{rgb}{0.3,0.3,0.7}
  \definecolor{light-blue}{rgb}{0.8,0.8,1}

\usepackage{hyperref}
\usepackage{setspace}
\hypersetup{
	colorlinks=true,
	linkcolor=dark-blue,
	citecolor=dark-red,
	urlcolor=dark-green,
	linktoc=page
}
\newcommand{\be}{\begin{equation}}
\newcommand{\ee}{\end{equation}}

\def\be{\begin{equation}}
\def\ee{\end{equation}}
\def\bea{\begin{eqnarray}}
\def\eea{\end{eqnarray}}

\newcommand{\rmd}{\mathrm{d}}

\newcommand{\w}{\wedge}

\newcommand{\f}[2]{\frac{#1}{#2}}
\renewcommand{\Im}{\text{Im}\,}
\renewcommand{\Re}{\text{Re}\,}

\newcommand{\e}{\textrm{e}}

\newcommand{\dd}{\mathrm{d}}

\DeclareMathOperator{\im}{Im}

\def\simleq{\; \raise0.3ex\hbox{$<$\kern-0.75em
      \raise-1.1ex\hbox{$\sim$}}\; }
   \def\simgeq{\; \raise0.3ex\hbox{$>$\kern-0.75em
      \raise-1.1ex\hbox{$\sim$}}\; }

\numberwithin{equation}{section}
\begin{document}

\begin{center}

{\LARGE {\bf The tension between 10D supergravity\\\vspace{0.4cm} and dS uplifts}}  \\

\vspace{1.5 cm} {\large  F. F. Gautason, V. Van Hemelryck, and T. Van Riet }\\
\vspace{0.5 cm}  \vspace{.15 cm} {Instituut voor Theoretische Fysica, K.U. Leuven,\\
Celestijnenlaan 200D B-3001 Leuven, Belgium
}

\vspace{0.7cm} {\small \upshape\ttfamily  ffg, vincent.vanhemelryck, thomas.vanriet @ kuleuven.be
 }  \\

\vspace{2cm}

{\bf Abstract}
\end{center}

{\small We elaborate on the recent work of Moritz et.~al. where it is argued that anti-brane uplifting in KKLT models never leads to dS vacua. This is due to flattening effects in the effective potential found when treating the gaugino condensate energy momentum in ten-dimensional supergravity.  We point out that the  Silverstein-Saltman uplift, which effectively dissolves the anti-brane in 3-form fluxes, is a conceptually simpler setting in which flattening effects can be present, without requiring assumptions about UV/IR mixings.  Along the way we revise and improve on the arguments of Moritz et.~al.~and discuss various subleties when studying gaugino condensates from a ten-dimensional point-of-view. In particular, the ``no dS'' argument is slightly weaker and we emphasize a technical loophole that might allow dS vacua.  Finally we suggest that AdS vacua which are parametrically controlled from these flattening effects belong to the Swampland. We comment on the relation with the recent refined dS Swampland inequalities.}

\setcounter{tocdepth}{2}
\newpage
\tableofcontents
\newpage

\section{Introduction}
The importance of establishing or disproving a dS landscape in string theory cannot be overstated. A landscape changes the way we think about fundamental physics and its relation to the constants of nature. If this landscape contains dS vacua with many possible sizes of the cosmological constant, then, in particular, the cosmic hierarchy problem might be solved by the anthropic principle \cite{Susskind:2003kw}.

However, the question of the consistency of dS vacua in string theory is not yet fully settled since none of the constructions are fully explicit and clear \cite{Danielsson:2018ztv}. Therefore there is still room for a conspiracy against dS solutions in string theory \cite{Brennan:2017rbf, Danielsson:2018ztv, Obied:2018sgi}\footnote{See also \cite{Bena:2017uuz}.}. There is more than one way forward in this debate. For instance, one can further scrutinize existing models. On the other hand, it would be interesting to come up with a general principle as to why dS vacua are impossible in quantum gravity. Perhaps this general principle is related to the possible inconsistency of the Bunch-Davies vacuum of quantum fields in dS space \cite{Danielsson:2018qpa}. In this paper, we pursue the first option, and, in particular, continue the scrutiny of KKLT-like models of dS space \cite{Kachru:2003aw} since these vacua form the prime examples that constitute evidence for a possible dS landscape.

As reviewed in \cite{Danielsson:2018ztv}, the KKLT scrutiny can roughly be divided into three categories. First, one could worry about the general principle to stabilise moduli by balancing quantum forces against classical forces \cite{Sethi:2017phn, Banks:2012hx}. This would already be a problem for the existence of the supersymmetric AdS KKLT landscape prior to uplifting. A second worry is the stability of the uplifting ingredient (the anti-brane), see for instance \cite{Bena:2009xk,  Gautason:2013zw, Michel:2014lva, Bena:2014jaa, Cohen-Maldonado:2015ssa, Bena:2016fqp,  Danielsson:2016cit} and references therein. Finally, the backreaction of moduli after supersymmetry-breaking should be carefully treated \cite{Moritz:2017xto}. This paper is about the latter. The main claim of \cite{Moritz:2017xto}\footnote{See \cite{Dasgupta:2014pma} for similar lines of thought.} is that anti-brane uplifting in KKLT models can lead to meta-stable AdS but never to dS. The reason is that the volume modulus shifts due to an interplay between the gaugino condensate and the supersymmetry-breaking source (the anti-brane) in such a way that only AdS is possible. In rough terms the problem is that the width of the KKLT potential, measured in terms of the mass of the volume modulus, $m^2_{\rho}$, is of the same order as the AdS cosmological constant:\footnote{To be more precise, in KKLT we have $m^2 L_{\text{AdS}}^2 = 4a^2\sigma_0^2$ where this is typically of order $10^3$. However it is never parametrically large.}
\begin{equation}
m^2_{\rho}\sim   \Lambda_{AdS} \rightarrow m^2 L_{\text{AdS}}^2 \sim 1\,.
\end{equation}
For an uplift to dS one needs to cancel the full negative energy, but then one looses control over how the volume modulus shifts. The effective potential cannot yet be computed directly in ten dimensions, but the sign of the resulting cosmological constant can. Reference \cite{Moritz:2017xto} found that it is never positive, indicating that the volume shift is significant. The ten-dimensional calculations should be consistent with four-dimensional effective field theory (EFT), see references \cite{deAlwis:2016cty, Kachru:2018aqn, Cicoli:2018kdo}. 

If the ten-dimensional computation of \cite{Moritz:2017xto} and its interpretation are correct, then the result has far-reaching consequences. However, there are some caveats to this story, which we list here.
\begin{enumerate}
\item The method of computation, which is briefly reviewed in this paper, relies on using a ten-dimensional description of the gaugino condensate following \cite{Dine:1985rz,Derendinger:1985kk,LopesCardoso:2003sp,Frey:2005zz,Baumann:2010sx,Baumann:2007ah,Dymarsky:2010mf,Heidenreich:2010ad,Koerber:2007xk}.  Obviously gaugino condensation is a strongly coupled IR 4D gauge theory effect that cannot be described classically. Still, one might hope that, as in \cite{Moritz:2017xto}, the way the gaugino condensate couples to gravity and other background fields can be captured semi-classically using the ten-dimensional description of 7-branes.
\item  The four-dimensional interpretation of the ten-dimensional computation implies that the anti-brane living in the IR of a warped throat, is influenced by the gaugino condensate living in the UV region, i.e.~out of the throat. This goes against naive field theory reasoning. An alternative viewpoint is that the gaugino strongly backreacts on the tip geometry giving an order one effect in the anti-brane energy \cite{Moritz:2017xto}.
\item  The computations in \cite{Moritz:2017xto} used several approximations on the backreacted solution.
\item  Finally, \cite{Moritz:2017xto} encountered diverging flux integrals in their computation and used an ambiguous regularization procedure based on a computational error.
\end{enumerate}
The aim of this paper is to discuss all of these issues to some extent. Concerning issue 2) we will demonstrate that a known alternative to anti-brane uplifting found by Silverstein and Saltman \cite{Saltman:2004sn} suffers the identical problem of \cite{Moritz:2017xto} but without involving UV/IR mixings. Issue 3) is addressed by revising the computation of \cite{Moritz:2017xto} and explicitly demonstrating that some approximations that were made can be dropped. Concerning issue 4) we correct the error in \cite{Moritz:2017xto} and observe that it slightly weakens their conclusions  and leaves an open door for dS uplifts.

The rest of this paper is organised as follows. In section \ref{sec:rev} we review the elements of anti-brane uplifting and the alternative flux uplifting of \cite{Saltman:2004sn} that we need for our discussion. In section \ref{sec:4D} we briefly review the problem of the flattening of the effective potential \cite{Moritz:2017xto} that is the source of thes tension with uplifting to de Sitter. This discussion will be within four-dimensional EFT. In Section \ref{sec:10D} we present and elaborate on the ten-dimensional computations underlying the idea of potential flattening and we touch upon the issues mentioned above. It will be made clear how the flux-uplifting scenario is captured automatically in this analysis. In section \ref{sec:CFT} we take a step back and question what is needed to have parametric control over dS uplifts. We will motivate that AdS vacua that lead to parametric control after uplift are likely to be in the Swampland. We end with a discussion in section \ref{sec:dis}. In Appendix \ref{App:10D} we give technical details on the ten-dimensional calculation presented in Section \ref{sec:10D}.

\section{Anti-brane and flux uplifting in KKLT models} \label{sec:rev}
In this section we briefly review KKLT moduli stabilisation \cite{Kachru:2003aw} and the difference between anti-brane uplifting \cite{Kachru:2003aw} and flux uplifting \cite{Saltman:2004sn}. This section serves as a review of a well-known story in order to make the paper self-contained. Experts are invited to jump directly to section \ref{sec:Backreaction}.

One of the long-standing problems with compactification of string theory down to four dimensions is the emergence of a plethora of massless scalar fields in four dimensions. These are directly related to the K{\"a}hler and complex structure deformations of the underlying compactification manifold which, in order to preserve supersymmetry, is a Calabi-Yau space. The first step in the KKLT program is to stabilise the complex structure moduli and the dilaton using three-form fluxes. This step is based on the DRS \cite{Dasgupta:1999ss} and GKP \cite{Giddings:2001yu} scenarios in type IIB string theory. Consider the ten-dimensional metric in Einstein frame
\begin{equation}\label{metric1}
\dd s^2_{10} = L^{-6} \e^{2A} \dd s^2_4 + L^2 \e^{-2A}\dd s^2_6 \,,
\end{equation}
where $L$ sets the overall volume of the compactification manifold with metric $\dd s_6^2$ and is one of the K{\"a}hler moduli. The warp factor, $\exp{A}$, is a function on the Calabi-Yau, but it can also posses $L$-dependence. Each pair of topological three-cycles on the Calabi-Yau space is associated with a complex structure modulus which can be stabilized by threading these cycles with NSNS and RR 3-form fluxes $H_3$ and $F_3$. These fluxes lead to a nontrivial profile for the self-dual 5-form and also require O3/D3-sources to cancel the RR tadpole. A direct dimensional reduction, ignoring the warp factor but taking into account the RR tadpole, leads to the following scalar potential:\footnote{In this paper we work in units where $2\pi\ell_s =1$. This choice means that the gravitational coupling constant is $1/(2\kappa_{10}^2) = 2\pi$.}
\begin{equation} \label{potential1}
V =  \f{2\pi}{L^{12}}\int\sqrt{g_6}~\f{\big| \star_6 G_3 - i G_3 \big|^2}{4\Im \tau} =\f{2\pi}{2L^{12}}\int\sqrt{g_6}~\e^{-\phi}\big|\star_6 H_3 + \e^{\phi} F_3\big|^2\,,
\end{equation}
where we have introduced the complex 3-form $G_3 = \dd C_2 - \tau \dd B_2$ and $\tau = C_0 +i \e^{-\phi}$.
Clearly the potential is a sum of squares, and a minimum is found when the 3-form flux obeys the Imaginary-Self-Dual (ISD) condition: 
\be \label{ISD}
\star_6 G_3 - i G_3 = 0~.
\ee
This condition stabilizes (some or all of) the complex structure moduli. However, in this vacuum  the radial mode $L$ is clearly massless. In fact, all K{\"a}hler moduli remain massless.

This setup can be recast in an ${\cal N}=1$ supergravity language in four dimensions. As such the scalar potential can be determined in terms of a K\"ahler potential $K$ and a superpotential $W$. To ease the presentation we specialize to a Calabi Yau space with a single K\"ahler modulus denoted by $\rho$. The relation between $\rho$ and $L$ is, $\sigma\equiv\Im\rho = L^4$, whereas the real part of $\rho$ is the axion one obtains by reducing $C_4$ over the unique 4-cycle. The complex structure moduli can be neatly packaged into the holomorphic 3-form $\Omega$ on the Calabi-Yau space and we have
\begin{align}
& W =  \int G_3\wedge \Omega\,,\\ 
& K ={ -3\log(2\Im\rho) - \log(2\Im\tau)-\log\Big(-i\int \Omega \wedge \bar{\Omega}\Big)}\label{Kaehler1}
\end{align}
Using this K{\"a}hler potential and superpotential one can easily verify that the scalar potential\footnote{The K\"ahler covariant derivative is defined by $D_a= \partial_a + \partial_a K$.}
\be\label{potentialintermsofKW}
V = e^K \left(G^{a\bar{b}}D_a W D_{\bar{b}} \overline W -3 W{\overline W}\right)\,,
\ee
simplifies to
\be\label{potential2}
V  = e^K G^{i\bar{\jmath}}(D_i W D_{\bar{\jmath}} W)\,,
\ee
where $G^{i\bar{\jmath}}$ is the inverse of the K{\"a}hler metric and the indices $i,\bar{\jmath}$ only run over the complex structure (CS) moduli and not the K{\"a}hler moduli.   As expected, this is a sum of squares, since it is nothing but a rewriting of our previous potential (\ref{potential1}) in an ${\cal N}=1$ language. The ISD condition (\ref{ISD}) naturally solves the F-term equations for the CS moduli $D_i W = 0$. Further imposing that ${\cal N}=1$ supersymmetry is preserved for these Minkowski vacua implies that $W=0$, which in turn implies that the ISD flux has no (0,3) piece. The same conclusion can of course be reached directly in ten dimension by studying the supersymmetry transformation of type IIB supergravity \cite{Grana:2000jj}.

It is important to note that the starting point of KKLT \cite{Kachru:2003aw} is one of the  supersymmetry-breaking ISD solutions described in the forgoing paragraphs. In particular, the complex structure moduli are stabilized at  a point for which $W$ is non-zero. Equivalently the ten-dimensional $G_3$ field has a non-vanishing (0,3) piece. The non-zero value for $W$  is denoted $W_0$  and since the complex structure moduli are already stabilized at high scale we will drop them from here on. Due to non-renormalization theorems that forbid perturbative corrections to the superpotential, the leading order correction to $W$ has to be non-perturbative and will lift the massless volume modulus:
\be\label{KKLT}
W = W_0 + {\cal A} \exp(i a \rho)\,. 
\ee 
This exponential term is argued to either arises from Euclidean D3 instantons or from gaugino condensation of the YM theory living on a stack of 7-brane wrapping the 4-cycle. We will discuss the latter in some detail in the next section. The parameters ${\cal A}$ and $a$ are determined by details of the origin of the non-perturbative effect. In particular $a= 2\pi/n$ with either $n=1$ for Euclidean D3-brane instantons or $2\pi/n$ is the dual Coexeter number of the gauge group living on the 7-branes in case of gaugino condensate. Within the framework of four-dimensional ${\cal N}=1$ supergravity the potential derived using the superpotential (\ref{KKLT}) and the original K{\"a}hler potential for the volume modulus in \eqref{Kaehler1} is
\be
V_\text{AdS}= \f{a\,\e^{-a\sigma}}{6\sigma^2}\left(3\Re(\overline{W_0}{\cal A}\,\e^{ia\theta}) + |{\cal A}|^2 \e^{-a\sigma}(3+a\sigma)\right)~,
\ee
where $\rho = \theta+i \sigma$. This potential admits a supersymmetric AdS vacuum with stabilised volume modulus.

\begin{figure}[h] \centering
\includegraphics[width=.6\textwidth]{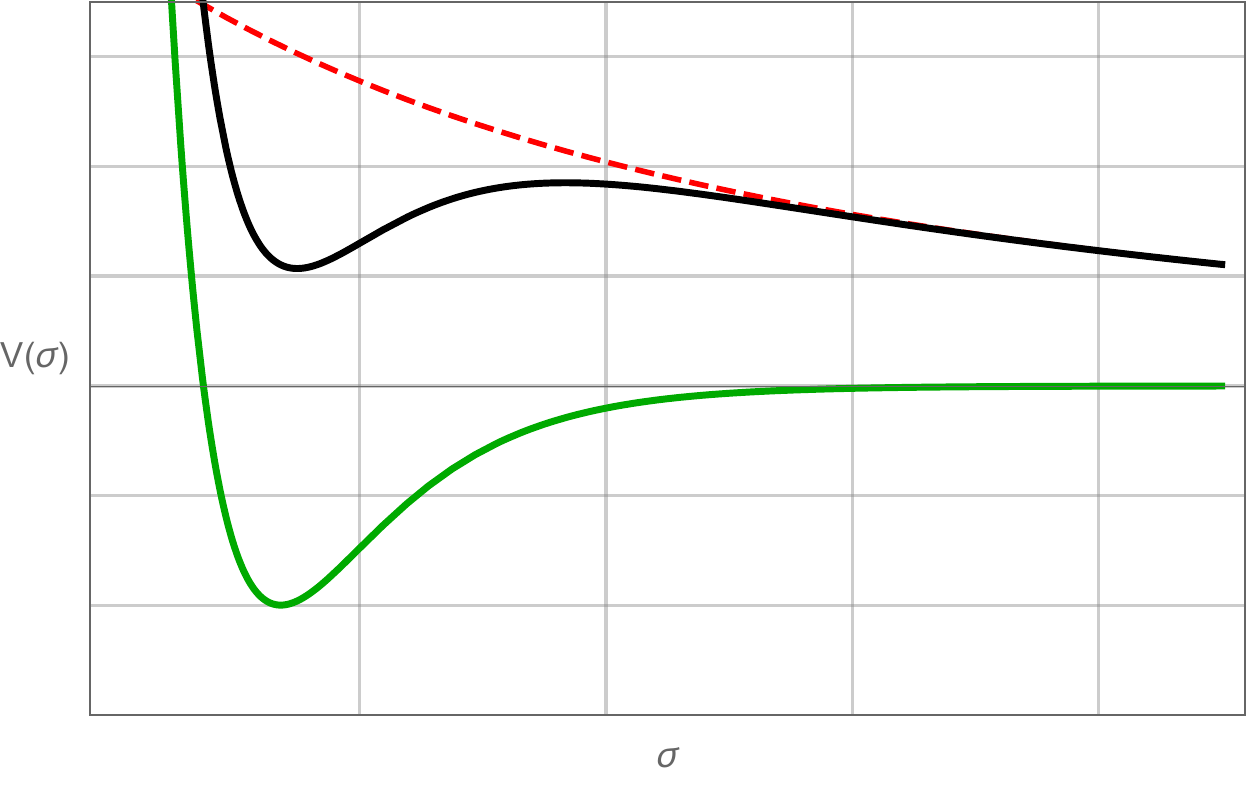}\caption{The KKLT potential: the sum of a powerlaw runaway uplift energy (red dashed line) with a sharply peaked AdS potential (solid green) from non-perturbative quantum effects can lead to meta-stable dS if the warp factor is suitably tuned (black solid curve). }%
\label{KKLTfig}%
\end{figure}

There are at least two proposals to modify the four-dimensional theory to lift this AdS vacuum into a dS meta-stable minimum \cite{Kachru:2003aw, Saltman:2004sn}. Originally the first proposal was to use anti-D3 branes. They preserve the opposite supercharges to the background and hence should contribute positive energy. Mobile anti-D3 branes are both gravitationally and electrically attracted towards regions of the Calabi-Yau space with the greatest warping. In particular, if the CY contains regions of extreme warping, so-called warped throats, their position is stabilized at the tip of the throat. Their string-scale energy is therefore drastically reduced to exponentially smaller values due to high warping. In a holographic context of non-compact throat geometries Kachru, Pearson and Verlinde (KPV) \cite{Kachru:2002gs} had already argued, using probe computations, that such anti-branes at the bottom of a warped throat are protected from direct brane-flux annihilation by making sure that the number of anti-branes is small compared to the 3-form flux number threading the three-cycle at the tip. In a compact setting, it is then argued that no new instabilities emerge since the redshift due to the warpfactor at the bottom of the throat can be dialed using the 3-form flux numbers. The total energy added by the branes is then twice their tension obtained by integrating (two times) the DBI term:
\begin{equation}\label{eq:antid3uplift}
V_{\text{uplift}} = \f{2\mu_3\, \e^{4A(\sigma)}}{\sigma^{3}} \approx  \f{2\mu_3\, \e^{4A}}{\sigma^{2}}\,,
\end{equation}
where $\mu_3$ is the tension of (anti-)D3 branes. It has been argued that the warp factor at the bottom of a throat depends  non-trivially  on $\sigma$ such that the overall $\sigma$-dependence in \eqref{eq:antid3uplift} is  $\sigma^{-2}$ as opposed to the naive expectation $\sigma^{-3}$ \cite{Kachru:2003sx}. The combined scalar potential contains the contribution from $W$ (\ref{KKLT}) together with the powerlaw uplift energy. The sum of the two nicely exhibits a meta-stable dS minimum as figure \ref{KKLTfig} demonstrates.
The two obvious decay channels are 1) tunnelling through the barrier displayed in the figure, which corresponds to a decompactification to ten dimensions and 2) the brane-flux decay of KPV where the anti-branes annihilate with the fluxes they are immersed in.

An alternative uplifting method to anti-branes was suggested by Saltman and Silverstein in \cite{Saltman:2004sn}. They observe that fluxes themselves could contribute positive energy just like anti-branes. We will henceforth refer to this as \emph{flux uplifting}. In many respects it does resemble the anti-brane uplift mechanism. In fact, flux uplifting can be thought of as anti-brane uplifting where the anti-branes have been dissolved into anti-imaginary-self-dual (AISD) 3-form fluxes. So the background contains both ISD fluxes to stabilize the complex structure moduli, and AISD fluxes to uplift. From our potential (\ref{potential1}) it is clear that whenever the flux is not purely ISD one finds a positive $\sigma^{-3}$ contribution to the potential energy. The question that arises then is whether one can consistently find a critical point, in the complex structure sector, away from the GKP vacuum. Unlike the GKP vacua, such critical points would not solve the classical ten-dimensional equations of motion, but if the non-perturbative correction is added, they would. Remarkably it was shown that one can stabilise the complex structure moduli away from the ISD point for a specific toroidal orientifold model without being plagued by tachyons \cite{Saltman:2004sn}. The example studied explicitly in \cite{Saltman:2004sn} most likely implies that many other Calabi-Yau manifolds exist for which similar results hold. A requirement for a phenomenological model is that the flux uplift term 
\begin{equation}
V_{\text{flux-uplift}} \propto  (\im\rho)^{-3}\int\sqrt{g_6}~\f{\big| \star_6 G_3 - i G_3 \big|^2}{4\Im \tau}\,,
\end{equation}
is small enough to be of the same order as the non-perturbative effects in \eqref{KKLT} and a dS critical point emerges.

\section{Backreaction in four and ten dimensions}\label{sec:Backreaction}

In this section we review a recent claim \cite{Moritz:2017xto} that the interaction of non-perturbative effects in \eqref{KKLT} and the uplift terms in \eqref{eq:antid3uplift} produce flattening effects that may spoil the dS critical point.  We separate the discussion into 4-dimensional effective field theory reasoning and ten-dimensional computations. In this section, we avoid discussing the technical aspects of the ten-dimensional treatment of gaugino condensation and postpone it until the appendix. The reason for this choice in presentation is to seperate the main idea put forth in \cite{Moritz:2017xto} from the computational details.

\subsection{Four-dimensional argument}\label{sec:4D}
The KKLT approach to uplifting stable AdS vacua to dS vacua is a general and intuitive idea. First, a stable (supersymmetric) AdS vacuum for which all scalar fields are perturbatively stable is identified. Then, a supersymmetry-breaking ingredient, whose dependence on the moduli is weak, is added to the construction. The moduli dependence must be weak enough so the stability of the AdS vacuum can be argued to extend to the dS vacuum, thereby also making this latter vacuum stable as well. For example, in the KKLT model, the AdS potential is sharply peaked, since it is derived from an exponential superpotential, whereas the anti-brane uplift energy follows a power law. 

Once supersymmetry is broken, it is reasonable to assume that the effective potential is not simply the sum of the supersymmetric AdS potential, and the supersymmetry-breaking ingredient. Typically we expect mixed terms that correct the naive sum of the two potentials:
\begin{equation}
V_{\text{total}} =  V_{\text{AdS}} + V_{\text{uplift}} + V_{\text{corrections}}\,.
\end{equation}
The details of the corrected potential depend on the scale of supersymmetry breaking as compared with the KK scale. Clearly, adding a  supersymmetry-breaking ingredient will lift the energy, but the correction terms could, in principle, destabilise the set-up. The total potential may therefore only allow for runaway solutions as opposed to meta-stable vacua. %In a typical uplift scenario this is especially important since the energy required to lift to dS space implies that one needs to add enough energy to bridge the negative cosmological constant.

A reasonable criterion for an AdS vacuum to be protected against such a runaway after uplift to dS would be that its potential well is parametrically narrow with respect to the energy gap that needs to be bridged by the supersymmetry-breaking ingredient. In other words, the ratio of the lightest scalar mass $m$ with respect to the cosmological constant needs to be large:
\begin{equation}
m^2L^2_\text{AdS}\gg 1\,.
\end{equation}
This condition is not fulfilled in KKLT models since that product is of order one. A possible way to have this inequality satisfied is to use the \emph{racetrack} superpotential \cite{Kallosh:2004yh}.  However, this method seems to violate the Weak Gravity Conjecture parametrically \cite{Moritz:2018sui}, so it is highly questionable whether the racetrack can be realized in a string theory compactification. {We will revisit this latter point in Section \ref{sec:CFT}}.

Returning to the KKLT set-up, we would like to derive, or at least estimate, $V_{\text{corrections}}$ in order to analyze the stability of the dS vacuum. It turns out that the formalism of non-linearly realized supersymmetry in four dimensions \cite{Volkov:1973ix, Komargodski:2009rz} is useful to this end, but it is not necessary. This approach limits the possible effective potentials $V_{\text{corrections}}$ one can write down and hence simplifies the discussion. The four-dimensional effective field theory of non-perturbative volume stabilization and anti-D3 brane supersymmetry-breaking within ${\cal N}=1$ supergravity was first developed in \cite{Kallosh:2014wsa, Bergshoeff:2015jxa, Kallosh:2015nia}. A nilpotent chiral superfield was essential when constructing these theories. It is still an open question whether the non-linear supersymmetry description is applicable to realistic compactification scenarios such as KKLT.\footnote{See \cite{Aalsma:2017ulu} for a more careful treatment and some open issues about the nilpotent description and \cite{Dudas:2016eej} for general critical remarks about its consistency. } Assuming the nilpotent description is valid, Moritz et.~al.~demonstrated in \cite{Moritz:2017xto} that it is powerful enough to constrain the form $V_{\text{corrections}}$. To understand this statement, let us start by first considering the standard KKLT potential in the framework of ${\cal N}=1$ supergravity with a nilpotent chiral superfield. The KKLT potential can be derived from the standard equation \eqref{potentialintermsofKW} using \cite{Ferrara:2014kva}
\be\begin{split}
& W(\rho, S) = W_0 + {\cal A} \e^{ia\rho} + \e^{2A}\sqrt{24 \mu_3} S\,,\\
& K = - 3 \log(2\Im\rho - S\bar{S})\label{Kahler1} \,,
\end{split}\ee 
where $S$ is the nilpotent field, $S^2=0$. The procedure to obtain the potential requires computing $V$ in the usual way using \eqref{potentialintermsofKW} and, at the end, setting $S=0$. This procedure leads to 
\begin{equation}
V = V_\text{AdS} + V_{\text{uplift}}\,.
\end{equation}

\begin{figure}[h] \centering
\includegraphics[width=.6\textwidth]{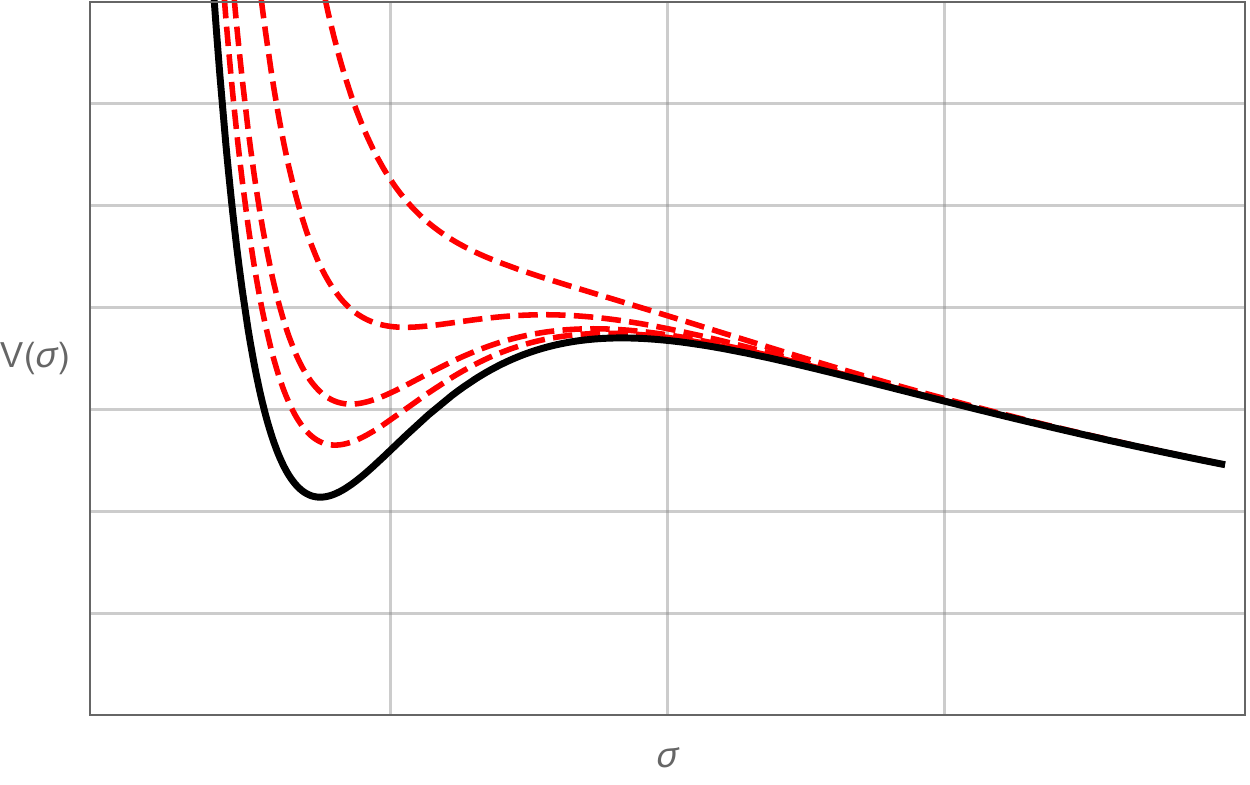}\caption{The KKLT potential plus correction term \eqref{Vcorr} resulting from the interplay between the anti-brane and the gaugino condensate. The black solid curve shows the undeformed KKLT potential which is valid for $c\sim\e^{2A}$. The red dashed curves show different values for $c$ ranging from $c \approx 0.1$ to $c\approx 1$. Other parameters in this plot are $W_0 = -10^{-4}$, ${\cal A}=1$, $a=0.1$ and $\e^{2A} \mu_4 = 2\times 10^{-11}$.}%
\label{DESYfig}%
\end{figure}

As explained in \cite{Moritz:2017xto} the nilpotent description is consistent with one extra term in $W$. This term is simply the product of the nilpotent field $S$ and the gaugino contribution $\e^{ia\rho}$. The physical interpretation of this term is the subtle interplay between the instanton determinant, ${\cal A}$, and  the supersymmetry-breaking effect captured in this formalism by $S$. Due to the nilpotency condition, the only dependence allowed is linear. The superpotential that should capture the interaction between the two effects can therefore be obtained by replacing
\begin{equation}
{\cal A} \rightarrow {\cal A}(1+cS)\,,
\end{equation}
in equation \eqref{Kahler1}. This term produces a correction term in the scalar potential, namely
\be\label{Vcorr}
V_{\text{corrections}}=\f{\e^{-a\sigma}}{12 \sigma^2}\left(2\sqrt{24\mu_3}~\Re\big({\cal A}c\,\e^{i a\theta}\big) + |{\cal A}c|^2\e^{-a\sigma}\right)\,.
\ee
In hindsight, the extra term in $W$ proportional to ${\cal A} c$ is not unexpected since when supersymmetry is not broken by anti-D3 branes, and some mobile D3 branes are present in the Calabi-Yau compactification, then it is known that ${\cal A}$ is a function of the D3 moduli. Hence, by pure analogy, ${\cal A}$ should also be a function of anti-D3 brane moduli. After integrating out heavy fields this dependence on the anti-D3 brane moduli produces a term ${\cal A} c S$.\footnote{The consistency of this extra term and its consequences for the KKLT model was debated in \cite{Kallosh:2018wme, Moritz:2018ani, Kallosh:2018psh}. If the outcome of this debate is that the coupling $cS$ should be suppressed in order to have a well-defined nilpotent description away from the vacuum, then we propose that the nilpotent description is not adequate to model the flattening effects.} The influence of this new term  on the stability of the vacuum is  determined by whether $c$ is warped down or not. If $c$ is warped down, i.e. it is proportional to $\e^{2A}$, it has negligible effects and the meta-stable KKLT dS vacuum is mostly unharmed. However if $c$ is of order one, the stability of the dS vacuum is affected and it can become perturbatively unstable as in fig. \ref{DESYfig} \cite{Moritz:2017xto}. 

It is important to note that in a compactification of string theory down to four dimensions, $c$ should not be thought of as a tunable number. In fact, it should be a function of the CY manifold the compactification is performed on, its complex structure moduli, and also, perhaps, the warp factor. Naively one would argue $c$ to be warped down since $c$ originates from  the anti-branes  located deep in the warped throat, and one expects that the anti-D3 backreaction remains confined in the throat. But, even if the anti-D3 backreaction is confined, this reasoning could fail. The backreaction of the gaugino condensate blueshifts towards the tip of the throat and could therefore have a significant effect on the tip geometry and thus the anti-branes. So having D7 branes in the UV could have an effect in the IR, and intuition based on non-compact throat geometries might be misleading. % Naively one could estimate this effect as $\langle \lambda\lambda \rangle \exp([-4A])$ and that number is order one if the uplift is to bridge the negative energy\footnote{Note that the negative energy is of order $\langle \lambda\lambda \rangle$ in string units and the uplift energy is order $p~ \e^{4A}$ with $p$ the number of anti-D3 branes. We thanks Jakob Moritz for explaining us.}. 
Whether or not $c$ is warped down is therefore a complicated and important issue. One way to settle it would be to provide a solid ten-dimensional argument as attempted in \cite{Moritz:2017xto}. The remainder of this section will be devoted to a review of the arguments of  \cite{Moritz:2017xto} and to point out subtleties with their analysis. % which ultimately invalidates some of their strong claims. This means that in our view the question of whether $c$ is warped down or not is still open.

\subsection{Off-shell vs on-shell potential}\label{sec:FreundRubin}
To appreciate the conclusions drawn from manipulating the ten-dimensional supergravity equations it is instructive to repeat some basic facts about compactifications. In particular the, perhaps confusing, difference between an off-shell and an on-shell potential obtained from dimensional reduction. For simplicity we illustrate the issue with a compactification to four-dimensions of a ten-dimensional theory with an arbitrary $p$-form field $F_p$ with $2\le p\le 6$ fixed. Our starting point is a simple ten-dimensional theory
\begin{equation}\label{10daction}
S = 2\pi\int \sqrt{|g_{10}|}\Bigl( R_{10} - \frac{1}{2}|F_p|^2\Bigr)\,.
\end{equation}
We now assume that the metric is a direct product metric of six-dimensional compact space and a four-dimensional spacetime
\be
\dd s_{10}^2 = \dd s_4^2 + \dd s_6^2\,.
\ee
We also assume that  the $F_p$-field has legs only along the 6 compact directions. With this assumption we want to compactify the theory down to a four-dimensional theory, obtaining an action of the form
\be\label{4daction}
S_4 = \f{1}{2\kappa_4^2}\int \sqrt{|g_{4}|}\Bigl( R_{4} -\f12 G_{ij}\partial_\mu \phi^i\partial^\mu \phi^j - V(\phi)\Bigr)\,.
\ee
where $\phi_i$ denotes arbitrarily many scalar fields obtained from the compactification, and $V$ is their potential. The four-dimensional gravitational constant is related to the ten-dimensional one through the standard relation
\be
\f{1}{2\kappa_4^2} = \f{1}{2\kappa_{10}^2}{\cal V}_6=2\pi{\cal V}_6\,,
\ee
where ${\cal V}_6$ is the volume of the compactification manifold. We use latin indices $m,n$ to denote ten-dimensional indices and greek indices $\mu,\nu$ as four-dimensional indices. It is clear from direct dimensional reduction of the action \eqref{10daction} that the fluxes contribute positively to the scalar potential:
\begin{equation}\label{fluxpot}
V_{\text{flux}} = \f{1}{2{\cal V}_6}\int \sqrt{g_6} \,|F_p|^2\,.
\end{equation}
 Despite this positive contribution, the  ten-dimensional Einstein equation can be used to obtain an apparently contradictory answer. First, consider the full trace-reversed Einstein equation
\begin{equation}
R_{mn} = \frac{1}{2}\frac{1}{(p-1)!} F_{mm_2\ldots m_p}F_{n}^{\phantom{n}m_2\ldots m_p} - \frac{p-1}{16}g_{mn}|F_p|^2\,,
\end{equation}
and evaluate it only over the four-dimensional indices
\begin{equation}\label{4Dtrace}
R_{\mu\nu} =  - \frac{(p-1)}{16} g_{\mu\nu}|F_{p}|^2\,.
\end{equation}
This should be compared with the four-dimensional Einstein equation derived from the general action \eqref{4daction} for a vacuum solution $\partial_\mu\phi^i=0$, which gives
\be
R_{\mu\nu} = \f12 g_{\mu\nu} V\,,
\ee
and so we can read of the potential. Surprisingly it is given by $V=-(p-1)|F_{p}|^2/8$ and is negative definite.\footnote{Our product manifold ansatz demands that $|F_p|^2$ is constant in the vacuum, therefore no integration is required when expressing the on-shell potential.} We should remember that this is not the full potential but rather it is the on-shell evaluation of the potential. This is consistent with the well-known Freund-Rubin solutions $AdS_4 \times X^6$ with $X_6$ a positively curved Einstein space, e.g. $S^6$. How can this be reconciled with the positive flux contribution to the scalar potential as in \eqref{fluxpot}? The trace-reversed Einstein equation has already incorporated the on-shell information that we assumed a static direct product metric Ansatz. If we would display explicit moduli dependence then the flux part of the scalar potential scales as $L^{-6-2p}$ and is hence runaway. There is no static solution. The only other contribution to the scalar potential comes from the curvature of the compact dimensions:
\begin{equation}
V_{\text{curvature}} =  - \int \sqrt{g_6}  R_6\,.
\end{equation}
If we would factor out the $L$-dependence, it would scale as $L^{-8}$. So again a negative power of $L$. Hence the only way to have a minimum of the sum $V_{\text{flux}}+V_{\text{curvature}}$ is if $R_6>0$ and the minimum is easily computed to be AdS. So the negative energy stored in the positive curvature of the compact dimensions, overshoots the positive energy of the fluxes. The size of the AdS curvature of course nicely corresponds to what can be found from equation (\ref{4Dtrace}). In other words, \emph{the ten-dimensional Einstein equation knew about the necessity of having positively curved compact dimensions in order to get a vacuum}. This is the crucial difference between on-shell and off-shell scalar potentials. When one obtains an expression for $R_4$ from manipulating ten-dimensional equations that include the trace-reversed Einstein equation, one finds the on-shell value for $V$. Whereas a direct dimensional reduction of the action gives the off-shell expression, but this is only useful if the moduli-dependence is known. In many cases it is not, In that case one can resort to the on-shell expression which at least tells us about the cosmological constant, but not much more. All information about stability, for example, is lost.  

Now that we have recalled these elementary facts, we are ready for a discussion about the KKLT vacuum from a ten-dimensional viewpoint.

\subsection{Ten-dimensional analysis}\label{sec:10D}
As we have reviewed, the AdS vacua of KKLT are build from the classical GKP solutions \cite{Dasgupta:1999ss,Giddings:2001yu} by adding the quantum effects that stabilise the volume modulus. In this paper we will focus on the effect originating from gaugino condensate on a stack of D7-branes. The field theory on a stack of D7-branes wrapping a four-cycle in the Calabi-Yau space descends to a ${\cal N}=1$ gauge theory sector of the four-dimensional effective theory. The Yang-Mills coupling constant of the gauge theory is determined by the volume of the four-cycle that the seven-branes wrap, $g_\text{YM}^2 = 4\pi/(g_s{\cal V}_4)$, where ${\cal V}_4$ is the warped volume of the four-cycle. For a single K{\"a}hler modulus Calabi-Yau the volume ${\cal V}_4$ is proportional to $\Im\rho$. The complexified gauge coupling of the field theory is similarly determined by the complex modulus $\rho$. Computing the parameter ${\cal A}$ is prohibitively complicated, but it should depend sensitively on the UV structure of the theory, and in particular the complex structure moduli. 
In order to establish the KKLT AdS (and later dS) vacua directly in ten dimensions requires a ten-dimensional understanding of the gaugino condensate. This may already sound counterintuitive since the gaugini condense in the IR, well below the KK scale. In \cite{Moritz:2017xto} it was argued, following earlier work \cite{Dine:1985rz, Derendinger:1985kk, LopesCardoso:2003sp, Frey:2005zz, Baumann:2010sx, Baumann:2007ah, Dymarsky:2010mf, Heidenreich:2010ad, Koerber:2007xk},that  one can work directly with the fermions living on probe seven-branes to capture the essential, leading order, physics of the problem. For example, one is supposed to be able to understand how the condensate gravitationally and electromagnetically with the other type IIB supergravity fields. It should be noted that this method has been both successful \cite{Baumann:2010sx,Baumann:2007ah} and unsuccessful \cite{Dine:1985rz,LopesCardoso:2003sp,Frey:2005zz} in reproducing the expected low-energy physics in the past. Therefore any conclusion obtained from this method should be taken with a grain of salt.

Keeping this in mind, we follow the above approach by including the fermion bilinear term in the effective action of seven-branes and compute its contribution to the ten-dimensional equations of motion that determine $R_4$. We first recall how this is done for the GKP solutions (and thus ignore quantum effects). Let us write the ten-dimensional metric as
\be
\rmd s_{10}^2 = e^{2A}\rmd s_4^2 + e^{-2A} \rmd s_6^2,\,.
\ee
and $C_4 = \alpha \star_4 1$ with $\alpha$, $A$ and the axion-dilaton as functions on the internal space only. Then a manipulation of the trace-reversed Einstein equation together with the $F_5$ Bianchi identity\footnote{In this manipulation one uses that the tadpole is canceled by negative tension sources (e.~g.~O3) that are BPS in the sense that their negative tension equals their charge.} leads to
\be\label{GKPnoqm}
-\dd \star_6\dd\Phi^-  = \star_6 \left(R_4 + \f{e^{2A}}{\Im\tau}\left|G_3^-\right|^2  + e^{-6A}|\dd \Phi^-|^2+\f{\e^{2A}}{2\pi} \Delta_\text{loc}\right)  \,.
\ee
where we have introduced the short-hand notation
\be\label{Deltadef}
\Phi^\pm = \e^{4A} \pm \alpha ~,\quad G_3^\pm = \f12\left(\star_6 G_3 \pm i G_3\right) ~,
\quad \Delta_\text{loc}=\f{1}{4}\left(T^m_m-T^\mu_\mu - 4\mu_3\rho_3\right)_\text{local}~.
\ee
The quantity $\Delta_\text{loc}$ denotes the contribution from localized sources and is computed by evaluating the energy-momentum tensors as well as total D3-brane charge density $\rho_3$ of localized objects in the geometry \cite{Giddings:2001yu}.
Note that even though this equation is a differential equation on the six-dimensional manifold alone, all metric contractions and hodge duals, are performed using the warped metric $\e^{-2A}\dd s_6^2$. %The only exception to this is the Hodge dual operator $\star_6$ which is performed using the unwarped metric. TVR: no the warpfactor cancels in a 6D hige start on a 3 form
Integrating this equation over the compact space and assuming that $\Delta_\text{loc}\ge0$ gives that the left hand side necessarily vanishes and that a Minkowsi vacuum is only possible if the 3-form $G_3$ is ISD and if $\Delta_\text{loc}=\Phi^-=0$ or equivalently $e^{4A}=\alpha$. This is the GKP solution. 

The localized sources available with $\Delta_\text{loc}=0$ are D3/D7 branes and O3 planes. The equation \eqref{GKPnoqm} cannot exclude the existence of AdS solutions with the ingredients used but obviously no dS solution can be obtained when $\Delta_\text{loc}\ge0$ since all terms on the right-hand side of \eqref{GKPnoqm} would be positive. 

The ten-dimensional approach to the KKLT AdS vacua is to add the extra ingredient on top of the GKP analysis, namely fermion bilinears living on localised sources (7-branes). We show in Appendix \ref{App:10D} how the contribution from localized sources now splits up into two contributions:
\be
\Delta_\text{loc} = \Delta_{\text{gaugino}} + \Delta_\text{BPS}~,
\ee
where $\Delta_\text{gaugino}$ is a contribution directly proportional to the fermion bilinear on the 7-branes $\langle \lambda\lambda\rangle$, and $\Delta_\text{BPS}$ is the classical contribution of D3, O3 and D7 branes as before, and therefore vanishes. Combining these results we can write (we display immediately the integrated result)
\be\label{eq:GKP_gauge}
{\cal V}_6 R_4 =-\int\star_6\left( \f{e^{2A}}{\Im\tau}\left| G_3^-\right|^2  + e^{-6A}|\dd \Phi^-|^2 + \f{\e^{2A}}{2\pi}\Delta_{\text{gaugino}}\right) \,.
\ee
For the discussion at hand, the form of $\Delta_{\text{gaugino}}$ is unimportant for now but will be explored in more detail in Appendix \ref{App:10D}. There, we will see that the term $\Delta_\text{gaugino}$ can be negative. In \cite{Moritz:2017xto} it was claimed that the combined term
\be
\f{\left|G_3^-\right|^2}{\Im\tau} + \f{\Delta_\text{gaugino}}{2\pi}~,
\ee
can be proven to be positive when reasonable assumptions are made. As it turns out this conclusion relied on a computational mistake in \cite{Moritz:2017xto}. In particular,  \cite{Moritz:2017xto} neglected to carefully treat divergences that arise due to the fermion bilinear. We will discuss these two issues in more detail in Appendix \ref{App:10D}. The bottom line is that we are unable to argue that the right-hand-side of \eqref{eq:GKP_gauge} has a definite negative sign. However, it is reasonable to expect that if the approach of treating gaugino condensate in ten dimensions is to make sense, then we must be able to recover the AdS$_4$ vacuum of KKLT from \eqref{eq:GKP_gauge}. The terms on the right-hand-side of \eqref{eq:GKP_gauge} must therefore combine in such a way as to overpower the positive contribution from $\Delta_\text{gaugino}$.  It is important to realize that the non-zero contribution to the integral in \eqref{eq:GKP_gauge} is concentrated in the region around the D7-branes since throughout the CY the solution is close to being GKP.

\subsection{How de Sitter could arise}\label{sec:dsworks?}

Let us now continue by adding the uplift ingredient, namely the anti-D3 brane. Again it can be accounted for by computing the combined contribution of localized sources to \eqref{GKPnoqm}
\be \label{fuckthatshit}
\Delta_\text{loc} = \Delta_{\text{gaugino}} + \Delta_{\overline{\text{D3}}} +\Delta_\text{BPS}= \Delta_{\text{gaugino}} + 2p\mu_3\delta_{\overline{\text{D3}}}~.
\ee
We presented the derivation of this contribution in Appendix \ref{App:10D}. The full integrated expression for the four-dimensional Ricci-scalar reduces to 
\be\label{punch}
{\cal V}_6 R_4 = - \int \star_6\left( \f{e^{2A}}{\Im\tau}\left| G_3^-\right|^2  + e^{-6A}|\dd \Phi^-|^2 + \f{\e^{2A}}{2\pi}\Delta_{\text{gaugino}} + \f{\e^{2A}}{2\pi}2p\mu_3\delta_{\overline{\text{D3}}}\right) \,,
\ee
where $p\mu_{3}$  \emph{is strictly positive}.  It might seem counterintuitive that the uplift ingredient, which is supposed to add energy, contributes with a negative sign to $R_4$. But the reason for this failed intuition is again that we are considering the on-shell potential. In fact, it has been shown that the GKP ingredients ($O3, F_3, H_3$) without gaugino condensation but with anti-D3 branes can lead to AdS solutions, but not dS solutions \cite{Blaback:2010sj}, and the negative contribution is consistent with that. The explanation is identical to our discussion for the Freund-Rubin solutions in Sec.~\ref{sec:FreundRubin}. Without a positive curvature of the internal dimensions the anti-D3 contribution would lead to a runaway. But the backreaction of the anti-D3 on the internal geometry indeed leads to a positive internal curvature whose negative energy overshoots the positive energy of the anti-branes. 

The only way a de Sitter vacuum is consistent with \eqref{punch} is that $\Delta_{\text{gaugino}}$ is now the dominant piece in the integral, given that this is the only positive contribution to the curvature.\footnote{This resonates well with suggestions that fermion condensate naturally lead to dS vacua. See \cite{Soueres:2017lmy, Terrisse:2018qjm} for recent papers on the matter.} This can happen if the backreaction of the anti-D3 branes on the geometry is such that all other terms in the expression \eqref{punch} are significantly reduced compared to their AdS counterpart. This scenario seems slightly contrived given our previous argument that in the AdS geometry a dominant piece of the integral arises close to the D7-branes since away from them the geometry is almost ISD. A backreaction that influences the terms around the D7-brane naively implies that the anti-D3 branes which are highly redshifted due to the large warp factor in the IR where they are sitting have a way to influence the UV geometry of the compactification where the D7-branes are. However if the backreaction occurs through the volume modulus as the four-dimensional effective theory suggests, then it is plausible that all terms in \eqref{punch} are affected in accordance with their volume scaling which can be complicated due to warping, see \cite{Junghans:2014xfa}. This ten-dimensional framework explicitly shows though that the anti-D3 brane must backreact rather strongly on the ten-dimensional geometry in order for the dS construction to work out.
It also explains how an uplift to non-supersymmetric AdS is possible where the vacuum energy should be less negative. This is completely consistent with the effect the anti-D3 brane has on the on-shell potential. As we explained, the influence on the cosmological constant is via the shifting of the volume to a larger value which dilutes the terms on the RHS of (\ref{punch}) to make them smaller. 

%This is clearly impossible since the assumption is exactly that the anti-D3 backreaction is confined to the IR of the warped throat whereas the gaugino condensate lives in the UV and gets not affected too much by the anti-D3. One could say it is exactly the demand that the anti-D3 is mildly bacreacting that causes the problem of not getting dS!. What certainly adds to the confusion is that from an EFT point of view this "non-backreaction" in fact implies the existence of an operator, $cS$ in the EFT that is sourced by the SUSY breaking that is nonetheless not warped down. But the right way to understand this is not as an anti-D3 backreaction effect that escaped from the throat to the UV, but rather a backreaction of the gaugino condensate that alters the tip of the throat where the anti-D3 sits. This is how the UV and the IR could ``communicate"\footnote{We thank Jakob Moritz for explaining this.}.

In the next subsection we discuss how very similar arguments can be applied to the flux-uplifting scenario reviewed towards the end of Sec. \ref{sec:rev}.  
%Before we proceed we would like to make one minor comment. 
%But what equation (\ref{punch}) tells us is that the flattening effect is too strong if one tries to uplift away all the negative energy. This simply does not work and the equation clearly indicates there is no static dS solution. That is why one expects a runaway instead if one keeps on adding anti-D3's to remove all the negative energy. Whether this runaway could lead to a quintessence model is not known to us. 

\subsection{Anti-brane vs flux-uplifting}

It should be clear that our ten-dimensional manipulations are not altered in any way when we turn to the flux-uplifting scenario. In fact, they rather simplify since now the anti D3-brane term is absent in equation (\ref{punch}). From a four-dimensional point of view this set-up is conceptually cleaner. One does not have to debate a certain operator being warped down or not, since the existence of the dS vacuum does not rest on a warped throat with a supersymmetry-breaking anti-D3 brane. Furthermore, one expects a flux flattening effect to be present. It is plausible that, after integrating out the complex structure scalars, the effective description of the flux-uplifting scenario can be written in terms of a nilpotent description. A priori one might care less about a nilpotent description since there is already a manifest $\mathcal{N}=1$ description before integrating out the complex structure scalars. But since we want to compare with anti-brane uplifting, and since the complex structure masses are expected to be heavier then the cut-off scale of our EFT, we integrate them out. The only difference with anti-brane uplifting is the scaling of the uplift term. It goes like $\sigma^{-3}$ instead of $\sigma^{-2}$. This can be captured by the following choice of $W$ and $K$:
\begin{align}
& W(\rho, S) = W_0 + \mathcal{A}\exp(ia\rho) + 2\sqrt{|G_-|^2} S\,,\nonumber\\
& K = - 3 \ln(2\Im\rho) - S\bar{S}\label{Kahler} \,,
\end{align} 
In this scenario it is quite hard to argue against writing an extra term in $W$ changing $\mathcal{A}$ to $\mathcal{A}(1+cS)$. After all, one expects the instanton determinant to be sensitive to the positions of the complex structure scalars. In anti-brane uplifting it is expected that the change in these positions is negligible after uplift. But for flux-uplifting the shift in the positions of these scalars is exactly what breaks supersymmetry and is supposed to generate dS vacua. So after integrating the CS scalars out, one is left with an $S$-dependence in the determinant $\mathcal{A}$, which can only be linear because $S^2=0$. But then one can easily loose the dS critical point if $c$ is not very small as shown in \cite{Moritz:2017xto}. From a field theory point of view there is no reason here why $c$ should be suppressed. 

What is less clear is whether one can see from the ten-dimensional computations (\ref{punch}) that $R_4$ is always negative. We left, on purpose, the term $\Delta_{\text{gaugino}}$ in (\ref{punch}) unspecified, aside from it being localised on a 4-surface inside the extra dimensions. It could be that this term is manifestly positive making $R_4$ manifest negative from equation (\ref{punch}). We do not pursue that option here. Instead we argue why (\ref{punch}) could imply $R_4 <0$ for the flux uplifting scenario, regardless of the details of $\Delta_{\text{gaugino}}$. Consider the opposite, namely that $\Delta_{\text{gaugino}}$ has a negative sign and can overcome the positive signs of the other terms in (\ref{punch}) such as to allow dS. That is logically possible, but it implies that the gaugino condensate has a drastic different energy contribution compared with its contribution in the supersymmetric vacuum, whereas in the effective potential analysis one had assumed that its form was roughly left untouched, apart from some mild change in volume. This argument is not watertight since we are reasoning from the point if view of an on-shell potential and one has to be careful in interpreting terms. Indeed the same caveats as for the KKLT model apply also here.

\section{Parametric control of uplift }\label{sec:CFT}
In section \ref{sec:4D} we argued that AdS$_4$ geometries that have all moduli stablized such that
\be\label{parametric}
m^2L^2_\text{AdS}\gg 1\,,
\ee
are robust against the flattening and destabilising effects caused by the uplift.  A main message of this paper is the suggestion that AdS vacua obeying (\ref{parametric}) parametrically are in the Swampland. Before we motivate this, we recall some basic facts about mass scales in KKLT-like vacua. 

In KKLT and related constructions, this criterion is not obviously satisfied for all scalar fields. The mass of the complex structure moduli is set by the volume of the internal manifold, $m_\text{CS} \approx 1/L^3$ \cite{Kachru:2003aw} where $L$ is the length scale set by the internal manifold as in \eqref{metric1} and is related to the volume modulus as before $\Im\rho=\sigma \approx L^4$.  First we note that the supersymmetric minimum of \eqref{KKLT} is located at $\rho=\rho_0 = \theta_0 + i \sigma_0$ determined by $D_\rho W=0$ which translates to
\be
3W_0 + {\cal A} \e^{i a\rho_0}(3+2a\sigma_0)=0~.
\ee
The AdS Length scale is determined by the potential evaluated at the minimum and is
\be
L_\text{AdS}^2 = \f{2}{\e^{K}|W|^2} = \f{36\,\e^{2a \sigma_0}\sigma_0}{a^2|{\cal A}|^2}\,.
\ee
For the volume modulus we must compute the mass of the canonically normalized scalar $z = \sqrt{6}\,(\rho-\rho_0)/2\sigma_0$ around the minimum and we find
\be
m_\rho^2  \approx \f{a^4|{\cal A}|^2\e^{-2a \sigma_0}\sigma_0}{9}\,,
\ee
where we used the leading order term in the large volume expansion for which $a\sigma_0\gg1$. We now see that
\be
m_\rho^2L_\text{AdS}^2 \approx 4a^2\sigma_0^2~.
\ee
This can be made big by tuning $a\sigma_0$ to be sufficiently large, in fact the non-perturbative term in \eqref{KKLT} is a first term in a series of terms that can only be truncated when $a\sigma_0\gg1$. It is clear that in KKLT one cannot satisfy (\ref{parametric}) \emph{parametrically} since $a\sigma_0$ cannot be tuned arbitrarily. 

Next let us consider the Kallosh-Linde (KL) model proposed in \cite{Kallosh:2004yh} where tuning is available. The KL model is based on the racetrack superpotential 
\be
W = W_0 + {\cal A} \exp(i a \rho) + {\cal B} \exp(i b \rho)\,,
\ee
where $b=2\pi/m$ with $m$ an integer. This model is argued to arise on two stacks of seven-branes wrapping the same homology 4-cycle but energetically separated from each other. In this case the vacuum is given by a solution of 
\be
3W_0 + {\cal A} \e^{i a\rho_0}(3+2a\sigma_0)+ {\cal B} \e^{i b\rho_0}(3+2b\sigma_0)=0\,.
\ee
This vacuum is AdS$_4$ with length scale
\be
L_\text{AdS}^2 = \f{36\sigma_0}{|a{\cal A}\e^{ia\rho_0} + b{\cal B}\e^{ib\rho_0} |^2}\,,
\ee
which can be dialed to arbitrary large values by carefully tuning the denominator to small values. In fact a Minkowski vacuum can be arranged by taking
\be
W_0 = -{\cal A}\left(\f{a{\cal A}}{b{\cal B}}\right)^{a/(b-a)} - {\cal B}\left(\f{a{\cal A}}{b{\cal B}}\right)^{b/(b-a)}~,
\ee
and still the volume modulus $\rho$ is stabilized at a finite value. The virtue of this model is that the mass of the lightest scalar $\rho$ can be made arbitrarily big when compared to the AdS length scale, and so the condition \eqref{parametric} is satisfied for all scalars. In  \cite{Moritz:2018sui} it was argued that exactly in this limit the model is in strong tension with the weak gravity conjecture which implies that the axion decay constant (in this case associated with the axionic partner $\theta$ of the volume modulus) cannot be parametrically large in Planck units. This is what motivates us to conjecture that AdS vacua with that parametrically satisfy (\ref{parametric}) are in the Swampland. 

To find more evidence for this conjecture it would be interesting to use holography to understand to what extent CFT principles might constrain the space of such AdS vacua.\footnote{We are grateful to Matt Buican for sharing notes about this idea.} %To that end, recall that by the above discussion, any modulus in such an AdS theory should have parametrically large mass, $m$, compared to the inverse AdS radius, $R^{-1}$
%\begin{equation}\label{massrel}
%mR=\kappa\gg1~.
%\end{equation}
When (\ref{parametric}) is satisfied for the lightest scalar field, AdS/CFT implies that the dual single-trace scalar operator has a parametrically large scaling dimension
\begin{equation}\label{dim}
\Delta={3\over2}+{1\over2}\sqrt{9+4(mL_{\text{AdS}})^2}\gg1~.
\end{equation}
Such theories may already seem somewhat contrived from the CFT perspective since they are \lq\lq dead-end" CFTs (i.e. they have no relevant deformations). For example, in 3D, there are no perturbative dead-end CFTs \cite{Nakayama:2015bwa}. However, there may be strongly coupled examples. We refer to \cite{Papadodimas:2011kn, deAlwis:2014wia, Polchinski:2009ch} for discussions about CFT duals to AdS landscape type vacua.

\section{Discussion} \label{sec:dis}

In this paper we have improved and elaborated on the results of \cite{Moritz:2017xto} concerning the validity of anti-brane uplifting in KKLT models. We have furthermore remarked that the results trivially extend to the flux-uplifting scenario of Silverstein and Saltman \cite{Saltman:2004sn}, where the tension between getting four-dimensional dS space and having a ten-dimensional treatment of the gaugino condensate is even sharper. One of our main results is the correction of a computational error in \cite{Moritz:2017xto} which makes the ``no-dS" result slightly weaker. As we have argued in section \ref{sec:dsworks?}, there could be room for a dS uplift.  Reasonable assumptions about the supergravity fields and the gaugino condensate make it seem unlikely that dS is possible after uplift. In fact the only loophole (within the formalism used) would be that the gaugino term in (\ref{punch}) provides the dominant energy after supersymmetry-breaking. This is not the case in the AdS vacuum and hence the anti-brane would have a rather drastic effect on the gaugino condensate energy. But as we emphasized there is little control about the actual form of that gaugino term in equation (\ref{punch}) especially since it probably diverges near the 7-brane source, where it needs a UV cut-off. It therefore seems to us that a way to make the gaugino condensate change drastically is by having an influence of the supersymmetry-breaking on the UV behavior, violating the assumptions of controlled supersymmetry-breaking in the IR. Another option is that the ten-dimensional description of gaugino condensation is flawed. The detailed description of this gaugino condensate does not affect the flattening as long as the condensate is a localised source of energy in ten dimensions.

As discussed in \cite{Moritz:2017xto}, a possible way around the flattening effects would be to find supersymmetric AdS vacua in string theory whose mass of the lightest mode $m^2$ is very high compared with the AdS mass scale $L_\text{AdS}^{-1}$:
\be\label{parametricdiscussion}
m^2 L_{\text{AdS}}^2 \gg 1\,.
\ee
If such AdS vacua exist then they are very gapped and their energy is only a bit negative (as compared with the scalar mass scale). Hence any small supersymmetry-breaking effect then necessarily leads to meta-stable dS vacua. We therefore conjecture, in light of the recent no dS conjectures, that such AdS vacua are in the Swampland. 

Indeed, one example of such AdS vacua can be found using KL racetrack fine-tuning \cite{Kallosh:2004yh}, but these models violate the Weak Gravity Conjecture parametrically \cite{Moritz:2018sui}. Another set of examples can be found using non-geometric fluxes as shown in \cite{Micu:2007rd, Palti:2007pm}. These examples are extreme cases of (\ref{parametricdiscussion}) because all moduli are fixed supersymmetrically in Minkowski space ($L_{\text{AdS}}\rightarrow \infty$). 
As we have explained these vacua can easily be turned into meta-stable dS vacua once supersymmetry is broken. Indeed, as shown in \cite{deCarlos:2009fq, deCarlos:2009qm, Blaback:2013ht} nearby these Minkowski vacua one can find meta-stable dS vacua. Our conjecture implies that these vacua, both the Minkowski and the de Sitter ones, are in the Swampland. This is perhaps not surprising since the constructions crucially rely on non-geometric fluxes, which are out of full control.

If the no-dS interpretation of the computations in  \cite{Moritz:2017xto} and this paper turn out correct then it fits with the recent conjecture that dS space cannot be found within string theory \cite{Danielsson:2018ztv, Brennan:2017rbf}. The fact that an uplift to meta-stable AdS is not ruled out seems in line of the even stronger swampland inequality \cite{Obied:2018sgi} 
\begin{equation}\label{Vafa1}
|\nabla V|\geq c V\,,
\end{equation}
with $c$ some positive order one number in Planck units. This inequality was further motivated in \cite{Ooguri:2018wrx, Hebecker:2018vxz}. It has become clear, due to the remarks in \cite{Denef:2018etk, Garg:2018reu, Conlon:2018eyr, Andriot:2018wzk} that this inequality should be used in case the Hessian is positive. In particular, \cite{Garg:2018reu, Ooguri:2018wrx}  refines the dS Swampland conjecture by stating that either ({\ref{Vafa1}}) or the minimum of the Hessian is bounded as follows:
\begin{equation}\label{Vafa2}
\text{min}{\nabla_i \partial_j V} \leq -c' V\,,
\end{equation} 
with $c'$ positive and of order one in Planck units.
Our conjecture that the lightest scalar in any AdS vacuum cannot satisfy (\ref{parametricdiscussion}) can equally rewritten as equation (\ref{Vafa2}) around critical points with negative or zero $V$.  

Finally, we note that our conjecture is also in mild tension with the standard KKLT AdS vacua, once one imposes the WGC for the axionic partner of the volume modulus. This follows from the relation between the axion decay constant and the mass of the volume modulus
\begin{equation}
m_{\rho}^2 L_{\text{AdS}}^2  \approx \frac{12}{f^2}\,,
\end{equation}
where $f$ is the decay constant in Planck units.

\section*{Acknowledgments}
We acknowledge useful discussions with  Nikolay Bobev, Arthur Hebecker, Shamit Kachru, Miguel Montero, Giotis M-theory notices Facebook page,  Eran Palti, Israel Ramirez, Matt Reece, Marco Scalisi, Gary Shiu, Cumrun Vafa, Timm Wrase and in particular Matthew Buican and Jakob Moritz for numereous discussions. FFG is a Postdoctoral Fellow of the Research Foundation
- Flanders. The work of TVR is supported by the FWO odysseus grant G.0.E52.14 and the C16/16/005 grant of the KULeuven. The work of VVH is supported by the by the European Research Council grant no.~ERC-2013-CoG 616732 HoloQosmos and the C16/16/005 grant of the KULeuven.  TVR gratefully acknowledges support from the Simons Center for Geometry and Physics, Stony Brook University at which some of the research for this paper was performed.

\appendix
\section{Gaugino condensation in 10D} \label{App:10D} 
\subsection{Conventions}
The type IIB supergravity action (in units where $2\pi \ell_s =1 $) is
\be
S = 2\pi\int\star_{10}\left(R_{10} - \f{\dd\tau \cdot\dd \bar\tau}{2(\Im\tau)^2} - \f{G_3\cdot \overline{G_3}}{2(\Im\tau)} - \f14 |F_5|^2\right) - \f{\pi}{2i}\int \f{C_4\w G_3\w \overline{G_3}}{\Im\tau}~,
\ee
where $R_{10}$ is the ten-dimensional Ricci scalar calculated using the metric $g_{MN}$ with determinant $g_{10}$. The axion-dilaton is denoted by $\tau = C_0+i\e^{-\phi}$ and the NSNS and RR 3-form field strengths have been combined into a single complex 3-form 
\be
G_3 = F_3 - i\e^{-\phi} H = \dd C_2 - \tau \dd B_2~,\quad H=\dd B_2~,\quad F_3 = \dd C_2 - H C_0~.
\ee
Throughout the paper we use short-hand notation to denote form contractions, let $\omega_p$ and $\psi_p$ denote two $p$-forms, then
\be
\star_{10}\omega_p \w \psi_p = \omega_p\cdot \psi_p~\star_{10}1~,\quad \omega_p\cdot \psi_p = \f{1}{p!}\omega_{M_1M_2\cdots M_p}\psi^{M_1M_2\cdots M_p}~,\quad |\omega_p|^2 = \omega_p\cdot\omega_p~.
\ee

\subsection{Anti-D3 contribution in equation (\ref{fuckthatshit})}
Before moving on to the more complicated seven-brane contribution to equation (\ref{fuckthatshit}) we first explain how quite counterintuitively the anti D3-brane provides a negative contribution to the four-dimensional vacuum energy. We start with the brane action of D3-branes with either positive or negative charge:
\be
S_{\text{D3}} = -p \mu_3 \int \sqrt{|P[g]}\dd^4 x  \pm p\mu_3 \int P[C_4]~,
\ee
where the upper sign refers to D3 and the lower to $\overline{\text{D3}}$, $p$ denotes the number of fundamental D3- or anti D3-branes and $P[\cdots]$ denotes the pull-back of corresponding fields to the brane worldvolume. The worldvolume of the D3-branes is taken to be the large space-time dimensions in order to preserve Lorentz invariance. In order to evaluate \eqref{Deltadef} we must compute the energy-momentum tensor contribution of the D3-branes as well as their charge density. These take the form
\be
T_{\mu\nu}^\text{D3} = -p\mu_3g_{\mu\nu}\delta_\text{D3}~,\qquad T_{mn}^\text{D3} = 0 ~,\qquad \rho_3^\text{D3} = \pm p\delta_\text{D3}~.
\ee
The combinations $\Delta_\text{D3}$ and $\Delta_{\overline{\text{D3}}}$ is now easily computed:
\be
\Delta_\text{D3} = 0~,\qquad \Delta_{\overline{\text{D3}}} = 2p\mu_3\delta_{\overline{\text{D3}}}~.
\ee

\subsection{Fermions on D7-branes}
As we discussed in  Section \ref{sec:10D} we will use the fermion bilinear on D7-branes to model the gaugino condensate responsible for the non-perturbative correction to $W$ in \eqref{KKLT}.  The bosonic action for such a brane in Einstein frame is given by:
\be\label{D7DBI}
S_{\text{D7}} = -\mu_7\int_{\Sigma_8} \dd^{8}x (\text{Im}\tau)^{-1}  \sqrt{|P[g]|}+\mu_7 \int_{\Sigma_8} P[C_8]~,
\ee
where $P[\cdots]$ denotes the pull-back of the corresponding fields to the brane worldvolume. This action should be supplemented with fermionic terms worked out in \cite{Camara:2004jj,Martucci:2005rb}. For our purposes the only non-zero fermionic contribution to the D7 action is given by:
\be\label{fermionaction}
S^{D7}_{\text{ferm}}=\int_{\mathcal{M}_{10}} \star_{10} \, (G_3 \cdot I + \overline{G_3}\cdot\overline{I})\,,
\ee
where 
\be \label{lambda2}
I = \pi \delta \frac{e^{-4A}}{\sqrt{\text{Im}\tau}} \frac{\bar{\lambda}\bar{\lambda}}{16 \pi^2}\Omega\,,
\ee
and $\delta$ is a delta function that localizes the the onto the eight-dimensional worldvolume. 
The worldvolume Majorana-Weyl spinor has been split into an internal and external part, their two Weyl-components being:
\be
    \theta_1 = \frac{1}{4\pi} e^{-2A} \lambda_D \otimes \eta + c.c. \quad \theta_2 = -\frac{i}{4\pi} e^{-2A} \lambda_D \otimes \eta + c.c. 
\ee
where $c.c.$ denotes charge conjugation. The external four-dimensional Weyl spinor $\lambda_D = (0,\bar{\lambda}^{\dot{\alpha}})^T$ gives rise to the fermion bilinear $\bar{\lambda^c_D}\lambda_D = -i\bar{\lambda}_{\dot{\alpha}}\bar{\lambda}^{\dot{\alpha}}$. For the remiander of this appendix we will denote the spinor bilinear $\bar{\lambda}_{\dot{\alpha}}\bar{\lambda}^{\dot{\alpha}}$ by simply  $\bar{\lambda}\bar{\lambda}$ and this is what appears in \eqref{lambda2}. The internal part of the spinor gives rise to the holomorphic three-form $\Omega_{ijk}=(\eta^{c})^\dagger \gamma_{ijk}\eta$. We assume that the fermion bilinear $\bar{\lambda}\bar{\lambda}$ acquires a non-vanishing expectation value in four-dimensional spacetime. We still denote this expectation value by $\bar{\lambda}\bar{\lambda}$ (and its complex conjugate by $\lambda\lambda$), which plays the role of the gaugino condensate.

\subsection{10D Einstein equation and subtleties }

The addition of the gaugino condensate to the four-dimensional superpotential, as in \eqref{KKLT}, admits a supersymmetric AdS solution. It is therefore a  useful consistency check to verify that the same conclusion can be reached by analysing the  on-shell potential \eqref{eq:GKP_gauge} when the contribution from \eqref{fermionaction} is included. For that one needs to use the equations of motion for the three-form fluxes and to compute the contribution $\Delta_\text{gaugino}$. This is computed to be \cite{thesis}
\be\label{eq:Delta_gauge_fermion}
	\Delta_\text{gaugino} = -\frac{3}{4}(G_3 \cdot I + \overline{G_3}\cdot\bar{I})
\ee
The bosonic action \eqref{D7DBI} does not contribute \eqref{eq:GKP_gauge} due to a BPS-type cancellation \cite{Giddings:2001yu}, so \eqref{eq:Delta_gauge_fermion} is the only contribution that the D7-brane delivers to the on-shell potential. 
%\begin{align}
%\label{eq:F3EoMgaugino}
%	&\dd \left(e^{\phi}e^{4A} \star_{6}F_3\right) + H \wedge \dd \alpha   = \frac{1}{2}\dd (e^{\phi}(X+\bar{X}))\\
%\label{eq:HEoMgaugino}
%	&\dd \left(e^{-\phi} e^{4A}\star_{6} H \right) +e^{\phi}e^{4A}\star_{6} F_3 \wedge F_1 + F_5 \wedge F_3
%	= -\frac{i}{2} \dd(\bar{X}-X)-\frac{1}{2}F_1 \wedge (e^{\phi}(X+ \bar{X})).
%\end{align}
Putting this last expression into \eqref{eq:GKP_gauge} does not immediately provide a definite sign for the scalar curvature $R_4$. The analysis of \cite{Moritz:2017xto} therefore reformulates \eqref{eq:Delta_gauge_fermion} by solving the $G_3$ equation of motion and plugging the result into \eqref{eq:Delta_gauge_fermion}. In what follows we review this strategy.

If we define the quantity $X$ to be 
\be
X = \delta \sqrt{\text{Im}\tau} \frac{\lambda\lambda}{16\pi^2} \star_6 \bar{\Omega} = \frac{i}{\pi}(\text{Im}\tau) e^{4A}\bar{I},
\ee
the $G_3$ equation of motion takes the nice form \cite{Moritz:2017xto}
\be
\label{eq:Lambda_EoM}
\dd \Lambda - \frac{\dd \tau}{\tau-\bar{\tau}} \wedge (\Lambda + \bar{\Lambda}) = \dd X - \frac{\dd \tau}{\tau-\bar{\tau}} \wedge (X + \bar{X}),
\ee
where 
\be
\Lambda = e^{4A}\star_6 G_3 - i \alpha G_3 = \Phi^+G_3^- + \Phi^- G_3^+\,.
\ee 
Then \cite{Moritz:2017xto} uses the Sen limit \cite{Sen:1996vd}, i.e. a configuration of four coincident D7-branes and a single O7-plane. In that case the DBI+CS action of the D7-brane cancels against the action of the O7-plane such that the type IIB axion and dilaton remain constant. Furthermore in the same limit the internal manifold is Ricci-flat. These facts can be verified directly from the equations of motion which read 
\begin{gather}
\label{eq:GKP_ISD_axion-dilaton_eq}
    \frac{1}{2}\frac{\nabla^2 \tau}{(\text{Im} \tau)^2} + \frac{i}{2}\frac{|\dd \tau|^2}{(\text{Im} \tau)^3} + \frac{1}{2\pi \sqrt{|g_{10}|}}\frac{\delta (S^{D7}+S^{O7})}{\delta \bar{\tau}} = 0,\\
\label{eq:GKP_ISD_internal_Einstein_eq}   
    \tilde{R}_{mn} = \frac{1}{2}\frac{\partial_{(m}\tau \partial_{n)}\bar{\tau}}{(\text{Im} \tau)^2} + \frac{1}{2}\left({T}^{D7}_{mn} - \frac{1}{8}{g}_{mn} {T}^{D7}\right)+ \frac{1}{2}\left({T}^{O7}_{mn} - \frac{1}{8}{g}_{mn} {T}^{O7}\right).
\end{gather}
Of course the gaugino condensate itself will induce a non-trivial coordinate dependence in $\tau$ and the internal curvature but these can be assumed to be constant to lowest order in the gaugino condensate. In the Sen limit the equation of motion \eqref{eq:Lambda_EoM}  reduces to just $\dd \Lambda = \dd X$. Next the authors of \cite{Moritz:2017xto} assumed that $\Phi^-$ vanishes such that the results of \cite{Baumann:2010sx} can be used to solve for $G_3$ to lowest order in terms of the gaugino condensate. The latter is done by  expressing the scalar delta function in terms of the holomorphic embedding equation $h$ of the four-cycle. The relevant expression is given by:
\be
\label{delta_function}
    \delta = \frac{1}{\pi} g^{m \bar{m}} \nabla_m \nabla_{\bar{m}} \text{Re} \log h.
\ee
The solutions for the $G_3$-flux are a (1,2) AISD and a (0,3) ISD part \cite{Baumann:2010sx}:
\begin{align}
\label{eq:G3min_solved}
    (e^{4A} G^{-}_3)_{k \bar{l} \bar{m}} &=  \frac{i}{2\pi}\sqrt{\text{Im}\tau}\frac{\lambda\lambda}{16\pi^2} \nabla_k \nabla_n (\text{Re} \log h) g^{n \bar{q}} \bar{\Omega}_{\bar{q}\bar{l}\bar{m}}\,,\\
 \label{eq:G3plus_solved}
    e^{4A} G^{+}_3 &= \frac{1}{2}X = \frac{i}{2\pi} \sqrt{\text{Im}\tau}\frac{\lambda\lambda}{16\pi^2} (g^{k \bar{l}} \nabla_k \nabla_{\bar{l}} (\text{Re} \log h)) \bar{\Omega}\,.
\end{align}
In these expressions we have corrected the expressions given in \cite{Moritz:2017xto} which were off by a  factor of $1/2$ in both equations. To obtain the solutions \eqref{eq:G3min_solved} and \eqref{eq:G3plus_solved} both a axion-dilaton as well as  the warp factor was assumed to be constant near the four-cycle to lowest order in $\lambda\lambda$.
Now the solutions \eqref{eq:G3min_solved} and \eqref{eq:G3plus_solved} can be substituted into \eqref{eq:Delta_gauge_fermion} to obtain
\be
	\Delta_\text{gaugino} = \frac{3i}{4}\left( G^+_3 \cdot I -  \overline{G^+_3} \cdot \bar{I}  \right)  = -\frac{3\pi}{4\text{Im}\tau} e^{-8A} |X|^2  = -\frac{3\pi}{\text{Im}\tau}|G^+_3|^2.
\ee
It is clear that this is manifestly negative. If this is substituted in the GKP equation \eqref{eq:GKP_gauge}, with $\Phi^- = 0$ as in \cite{Moritz:2017xto}, we find
\be
  	\mathcal{V}_6 R_4 = -\int \star_6 \frac{e^{2A}}{\text{Im}\tau}\left(| G^-_3|^2 -\frac{3}{2} |G^+_3|^2 \right).
\ee
If we substitute the solutions \eqref{eq:G3min_solved}-\eqref{eq:G3plus_solved} we obtain
\begin{align}
\label{eq:GKP_gauge_fermion}
\begin{split}
  	-\mathcal{V}_6 R_4 =& -\frac{2}{\pi^2}\int \star_6 \: e^{-8A}\left| \frac{\lambda\lambda}{16\pi^2}\right|^2 (\nabla_k \nabla_l (\text{Re} \log h))(\nabla^k \nabla^l (\text{Re} \log h))\\
  	& +\frac{3}{\pi^2}\int \star_6 \: e^{-8A}\left| \frac{\lambda\lambda}{16\pi^2}\right|^2 |g^{k \bar{l}} \nabla_k \nabla_{\bar{l}} (\text{Re} \log h)|^2.
\end{split}
\end{align}
The authors of \cite{Moritz:2017xto} claim that the last term can be integrated by parts such that it has the same integrant as the first term. With their prefactors (8 and 6) it is consitent with an AdS spacetime. With the corrected factors however, this argument is no longer valid. There are some subtleties that arise, since both integrals in \eqref{eq:GKP_gauge_fermion} are divergent in the UV, with different singular behaviour of both integrants\footnote{This was suggested to us by Jakob Moritz.}. Appropriate UV-regularisation of both integrals choosing a correct cut-off should be able to make the right hand side positive, giving evidence for an AdS spacetime. We do however not provide the correct regularisation procedure in this paper.

Another procedure hints to the same conclusions. The strategy above was to calculate the source term $\Delta_\text{gaugino}$ using solutions to the equations of motion, expressing $\Delta_\text{gaugino}$ in terms of the delta function $\delta$. Another strategy tries to avoid this using the three-form flux equations of motion to rewrite \eqref{eq:Delta_gauge_fermion} up to a total derivative. For the details of this calculation we refer to \cite{thesis}. The end-result is given by 
\be
\label{eq:Delta_gaugino_Vincent}
    e^{-2A}\Delta_\text{gaugino} \; \star_{6} 1 = -\frac{3\pi e^{-6A}}{2\text{Im}~\tau} \left(\Phi^+ |G_3^{-}|^2 \ + \Phi^- |G^+_3|^2\right) \star_6 1 + \text{total derivative}\,.
\ee
This result requires no approximations, in contrast to the previous approach. If we assume that the warp factor and axion-dilaton are constant and $\Phi^- = 0$, as in \cite{Moritz:2017xto}, then we can substitute \eqref{eq:Delta_gaugino_Vincent} into \eqref{eq:GKP_gauge} to find
\be
  	\mathcal{V}_6 R_4 = +\frac{1}{2}\int \star_6 \: \frac{e^{2A}}{\text{Im}\tau}| G^-_3|^2 - \int  \left(\star_6 ~e^{4A} \cdot \text{total derivative}\right).
\ee
Since the warp factor is constant, the last integral should vanish if singularities are absent. If so, the leftover contribution at the right hand side forbids AdS, in contradiction to expectations. We see several possible resolutions of apparent puzzle. Most prominently we believe the approximations by \cite{Moritz:2017xto} are much too restrictive, especially the constant warping and vanishing of $\Phi^-$, which clearly does not hold near sources. This would imply that the ``total derivative'' term is not a total derivative. Secondly, since the fluxes can be singular near the brane, integrals might need regularization. The `total derivative' that acts on fluxes then does not vanish upon integration. Another possibility is that there is no ten-dimensional description of KKLT AdS vacua.

\bibliographystyle{utphys}

    \bibliography{refs}

\end{document}